# Magnetic Imaging of Macroscopic Spin Chirality Flipping

H. Miao[1], G. Fabbris[2], J. Bouaziz[3], W. R. Meier[4], P. Mercado Lozano[2], Y. Choi[2], J. Strempfer[2], D. Haskel[2], S. Blügel[3], M. Cook[1], M. Brahlek[1], H. N. Lee[1], A. D. Christianson[1], A. F. May[1], S. Okamoto[1]

[1]Materials Science and Technology Division, Oak Ridge National Laboratory, Oak Ridge, Tennessee 37831, USA

[2]Advanced Photon Source, Argonne National Laboratory, Argonne, Illinois 60439, USA

[3]Peter Grünberg Institut and Institute for Advanced Simulation, Forschungszentrum Jülich and JARA, D-52425 Jülich, Germany

[4]Department of Materials Science and Engineering, University of Tennessee, Knoxville, Tennessee, USA

**Chirality is a fundamental organizing principle of correlated and topological states.[1-4] In quantum magnets, chirality arises from the geometric twisting of spins and serves as an emergent source of Berry curvature and quantum metrics.[5-11] Although external fields can reversibly tune the spin chirality, understanding how spontaneous reversal occurs on macroscopic length scale remains an unresolved challenge. In this letter, we use resonant magnetic x-ray scattering with 2.5-μm spatial resolution to image intertwined spin, charge, and lattice orders of the correlated topological magnet EuAl$_4$.[12-19] We uncover a macroscopic chirality flipping transition and a remarkable chiral memory effect. The chiral magnetic domain tracks the landscape of the underlying charge density wave, implicating emergent chiral magnetic interactions arising from competing chiral and nematic lattice fields. Our results reveal the fundamental significance of magnetoelastic coupling in stabilizing homochiral and topological magnetic states.**

Strongly correlated electronic systems are fertile platforms for the emergence of macroscopic quantum states, including high-$T_c$ superconductivity and quantum spin liquids.[4] In these systems, many-body interactions involving spin, charge, orbital, and lattice degrees of freedom are intertwined, giving rise to nearly degenerate electronic instabilities that are sensitive to subtle variations in local charge density and lattice distortions. Uncovering the real-space organization of correlated states and their responses to thermal and quantum fluctuations are essential for deciphering their underlying microscopic mechanisms.

Chiral magnets represent a prototypical class of correlated states, exhibiting rich fundamental physics and strong potential for quantum and spintronic applications.[5-11] In these systems, the spin chirality, $\chi_s$=1 or -1, can be inherited from the inversion-symmetry breaking lattice structures, through the relativistic Dzyaloshinskii-Moriya (DM) interaction,[20,21] or emerge spontaneously via competing magnetic interactions, such as the real-space oscillating Ruderman-Kittel-Kasuya-Yosida (RKKY) interaction.[8,22-26] Flipping of $\chi_s$ generally requires external fields, such as electromagnetic and strain field. Interestingly, recent experimental studies[15,16] have revealed an spontaneous $\chi_s$-flipping transition in the chiral spin and charge ordered EuAl$_4$, raising questions on the interplay between spin, charge, and lattice orders, particularly the principles governing their real-space organization. In this letter, using x-ray resonant magnetic scattering (XRMS) with 2.5 μm spatial resolution (Fig. 1a), we uncover the microscopic signature of macroscopic $\chi_s$-flipping transition that is driven by competing chiral and nematic lattice field in EuAl$_4$.

EuAl$_4$ hosts a cascade of symmetry breaking transitions upon cooling, culminating with a final transition marked by the spontaneous $\chi_s$-flipping as shown in Fig.1b.[12-19] At room temperature, EuAl$_4$ is centrosymmetric with *I4/mmm* symmetry. Below $T_{cdw}$=140 K, the material experiences a transverse Peierls transition[18] resulting in a charge density wave (CDW) with a wavevector $\boldsymbol{q}_{cdw}$=(0, 0, 0.183) in reciprocal lattice unit (r.l.u.) along the crystalline c-axis. The corresponding lattice deformation involves primarily Al displacements in the ab-plane that break inversion and mirror-symmetries but preserves (screw) fourfold rotation-symmetry, C4.[17-19] At $T_1$=15.4 K, the system breaks time-reversal symmetry, $\mathcal{T}$, and forms a double-Q spin density wave (SDW) with the spin moment confined in the ab-plane. A spin flip transition happens at $T_2$=13.3 K, where the spin moment gains a finite c-axis component.[13-15] The C4 lattice symmetry breaks at $T_{nem}$=12.3 K, yielding a chiral SDW and negative helicity with $\boldsymbol{q}_\chi^-$=(0.165, 0, 0). Most interestingly, at $T_\chi$=9.8 K, the material experiences a $\chi_s$-flipping transition resulting in a chiral SDW with a slightly larger $\boldsymbol{q}_\chi^+$=(0.185, 0, 0).[15,16]

To reveal the nature of the $\chi_s$-flipping transition, we use XRMS at the Eu $L_2$-edges, corresponding to the resonant $2p_{1/2} \rightarrow 5d$ transitions shown in Fig. 1a. The resonant scattering amplitude from site *n* can be written as[27] (see also Methods):

$$f_n = (\hat{\epsilon}' \cdot \hat{\epsilon})f_0 - i(\hat{\epsilon}' \times \hat{\epsilon}) \cdot \widehat{M}_n f_1 \quad (1)$$

Where $\hat{\epsilon}$ and $\hat{\epsilon}'$ are the polarization vectors of the incident and scattered x-rays, respectively. $\widehat{M}_n$ is the direction of the magnetic moment of site $n$. $f_0$ and $f_1$ are proportional to the anomalous charge scattering form factor and the linear magnetic scattering form factor, respectively. For the uniaxial chiral SDW, such as the $\boldsymbol{q}_\chi^-$ and $\boldsymbol{q}_\chi^+$ phases, the magnetic scattering in the rocking geometry (see Fig. 1a) displays giant circular dichroism (CD) that is directly related to the $\chi_s$[28]:

$$I^{CR}(\boldsymbol{Q}_\chi^\pm) - I^{CL}(\boldsymbol{Q}_\chi^\pm) = (\tau\chi_s)\mathcal{D}^{xz}(\boldsymbol{Q}_\chi^\pm) \quad (2)$$

where $\tau = sign((\boldsymbol{Q}_\chi^\pm \cdot \hat{x})/|\boldsymbol{Q}_\chi^\pm \cdot \hat{x}|)$. The $\hat{x}$ is the unit vector along the x-direction (Fig. 1a) and $\boldsymbol{Q}_\chi^\pm = \boldsymbol{q}_\chi^\pm + \boldsymbol{G}$ is the magnetic superlattice peak position in Brillouin zone $\boldsymbol{G}$. The $\mathcal{D}^{xz}$ is a function of magnetic moment that is independent of $\tau$ and $\chi_s$. CR and CL stand for circular right and circular left incident photon polarization, respectively. Combining XRMS and high lateral spatial resolution (see Methods), one can map the real-space intensity distribution, $I^{\hat{\epsilon}}(\boldsymbol{Q}, x, y)$. The momentum transfer, $\boldsymbol{Q} = \boldsymbol{k}_f - \boldsymbol{k}_i$, can be tuned to the $\boldsymbol{Q}_\chi^\pm$, $\boldsymbol{Q}_{cdw}$, and $\boldsymbol{G}$ to selectively probe chiral spin, charge, and lattice orders. Figure 1c and 1d show prototypical $I^{CR}(\boldsymbol{Q}_\chi^+, x, y)$ and $I^{CL}(\boldsymbol{Q}_\chi^+, x, y)$ maps at 8.5 K that reveal SDW domains. These maps show a large circular polarization dependence, in agreement with the chiral SDW order.[15]

**Macroscopic $\chi_s$-flipping and chiral memory effect**

We first establish the $\chi_s$-flipping across $T_\chi$. Figure 2a shows the CD-XRMS imaging, $R^{CD}(\boldsymbol{Q}, x, y) = \frac{I^{CR}(\boldsymbol{Q},x,y) - I^{CL}(\boldsymbol{Q},x,y)}{I^{CR}(\boldsymbol{Q},x,y) + I^{CL}(\boldsymbol{Q},x,y)}$, at $T$=10.2 K and $\boldsymbol{Q} = \boldsymbol{Q}_\chi^- = (0.165, 0, 6)$. We find that in the 400×400 μm$^2$ scanning area, the sample is dominated by one SDW domain. The large negative CD in our experimental geometry reveals $\chi_s = -1$. Figure 2b shows the CD-XRMS image in the same field of view (FOV) but at $T$=8.5 K and $\boldsymbol{Q} = \boldsymbol{Q}_\chi^+ = (0.185, 0, 6)$ (*see Supplementary Fig.S1 for CD-XRMS images at $\pm\boldsymbol{Q}_\chi^\pm$*). We find that the negative intensity of the entire chiral domain changes to positive, establishing a spontaneous $\chi_s$-flipping transition on a macroscopic scale.

To further understand the self-organization of the macroscopic chiral domain, the sample was warmed up to the paramagnetic phase at $T$= 20 K and then cooled back to the chiral magnetic phases. The CD-XRMS images at $T$=10.2 K and 8.5 K after thermal cycling are shown in Figs. 2**c** and **d**, respectively. Remarkable, the large chiral domain, including its shape, $\chi_s$ and $\chi_s$-flipping behavior, remain the same, representing a chiral memory effect. We, however, note that a relatively smaller chiral magnetic domain on the top left corner retains a negative $\chi_s$ below $T_\chi$, possibly due to the presence of small defects such as strained regions (*see Supplementary Fig.S2*).

**The coupling between $\chi_s$ and $q_\chi^\pm$**

Since $q_{SDW}$ changes discontinuously from $q_\chi^- = (0.165, 0, 0)$ to $q_\chi^+ = (0.185, 0, 0)$ at $T_\chi$, this naturally raises the question of whether the chiral magnetic interaction is $q$-dependent. To understand this scenario, we determine the temperature dependence of CD-XRMS line-scans along the (H, 0, 6) direction. Fig. 3**a** shows these line scans that were measured on the dashed white square in Fig. 2**c**. We observe a $\Delta T$~1K temperature window where $q_\chi^-$ and $q_\chi^+$ coexist, in agreement with a first order phase transition. Surprisingly, in this regime of phase coexistence, $q_\chi^-$ and $q_\chi^+$ phases both show positive $\chi_s$ (*see Supplementary Fig. S3 for the CD-XRMS image*). The $\chi_s$-flipping appears only after the $q_\chi^+$ phase is suppressed completely. These observations exclude mechanisms based on a $q$-dependence of chiral magnetic interactions and strongly point to hidden local field that drives both $\chi_s$-flipping and $q_\chi^+$-phase.

**CDW pinning of chiral magnetic domain**

As shown in Fig. 1**b**, the $\chi_s$-flipping transition develops within the CDW and C2 structural phases, so it is plausible that the $\chi_s$ couples to local chiral and nematic fields.[18] To reveal this interplay, we image SDW, CDW, and structural domains in the same FOV. Figures 4**a** and 4**b** show $I^{CR}(Q, x, y)$ at $Q_{a,\chi}^+$=(1.189, 0, 7) and $Q_{b,\chi}^+$=(1, 0.189, 7) at $T$=5.5 K. These images correspond to orthogonal SDW domains with spin density modulations along the crystalline a-axis and b-axis, respectively (*see Supplementary Fig.S4 and Fig. S5 for more images*). Figures 4**c** and 4**d** show $I^{CR}(Q, x, y)$ at $Q_{CDW}$=(1, 0, 7.167) and $G$=(1, 0, 7) that respectively image the CDW and principal structural domains. As a reference, $I^{CR}(G, x, y)$ shows a large and homogenous domain at the center of the image. In contrast, both SDW and CDW show large intensity variations within the big structural domain. For instance, we see clearly that the $I^{CR}(\mathbf{Q}_{a,\chi}^+, x, y)$ (Fig. 4a) and $I^{CR}(\mathbf{Q}_{b,\chi}^+, x, y)$ (Fig. 4b) are spatially separated, as expected for orthogonal domains. Most interestingly, we find that $I^{CR}(\mathbf{Q}_{a,\chi}^+, x, y)$ is nearly identical to $I^{CR}(\mathbf{Q}_{CDW}, x, y)$ (Fig. 4c) and showing the same diagonal line gaps (highlighted by red dashed lines) that is filled by the $I^{CR}(\mathbf{Q}_{b,\chi}^+, x, y)$. These results strongly support the CDW pinning of chiral magnetic domains.

**Competing chiral and nematic fields and $\chi_s$-flipping**

It has been shown that the magnetic interactions in EuAl$_4$ are primarily determined by the long-range RKKY interaction.[15] The observations of a chiral memory effect and CDW pinning of chiral magnetic domain indicate emergent chiral magnetic correlations that are locally competing. In the paramagnetic phase, since CDW breaks inversion and mirror-symmetries but preserves (screw) C4-symmetry, it gives rise to an effective chiral lattice field, $\boldsymbol{\psi}_{chi}$, that is proportional to an out-of-plane polarization vector, $\boldsymbol{P}_1$, as depicted in Fig. 4e. This chiral structure will give rise to a chiral magnetic interaction, $\boldsymbol{D}_1 = \chi_- \boldsymbol{\psi}_{chi} \times \boldsymbol{r}_{12}$, where $\boldsymbol{r}_{12}$ links nearest magnetic atoms. Below the (screw) C4 to C2 transition at $T_{nem}$, the lattice field is expected to have enhanced nematic field, $\boldsymbol{\psi}_{nem}$, that is proportional to an in-plane polarization, $\boldsymbol{P}_2$, as schematically shown in Fig. 4g. This change induces a new chiral magnetic interaction, $\boldsymbol{D}_2 = \chi_+ \boldsymbol{\psi}_{nem}$. We note that $\boldsymbol{D}_2$, $\boldsymbol{\psi}_{nem}$, and $\boldsymbol{q}_\chi$ are parallel vectors as required by the symmetry and helical spin texture. As shown in Figs. 4f and h, while $\boldsymbol{\psi}_{chi}$ and $\boldsymbol{\psi}_{nem}$ can locally coexist, they are competing orders, because $\boldsymbol{\psi}_{chi}$ favors C4-symmetry and an *s*-wave-like CDW gap in the electronic structure, whereas $\boldsymbol{\psi}_{nem}$ breaks C4-symmetry and host a *p*-wave-like CDW gap in the electronic structure.[18] The competition between $\boldsymbol{\psi}_{chi}$ and $\boldsymbol{\psi}_{nem}$ results in competing chiral magnetic interactions and drives the spontaneous $\chi_s$-flipping transition. In this picture, the $\boldsymbol{q}_\chi^-$ phase is a metastable phase arising from the microscopic coexistence of $\boldsymbol{\psi}_{chi}$ and $\boldsymbol{\psi}_{nem}$ as shown in Fig. 4f. Indeed, the $\chi_s$-flipping transition coincides with the commensurate-to-incommensurate CDW transition[15], compatible with competing lattice fields. Furthermore, x-ray scattering has shown that $\boldsymbol{\psi}_{nem}$ increases monotonically with cooling below $T_{nem}$[18], consistent with monotonic enhancement of $\boldsymbol{D}_2$. It is important to emphasize that the CDW induced lattice distortions in EuAl$_4$ are on the picometer length scale.[17-19] The sensitivity of $\chi_s$ to small charge and lattice fields highlights the fundamental importance of magnetoelastic coupling and hence critical for the understanding of growing number of nanometric skyrmion platforms.[13,22-26]

In summary, using XRMS, we imaged the temperature dependent evolution of chiral SDW domains. We uncovered a chiral memory effect and established the CDW pinning of chiral magnetic domains. Our results uncover the fundamental significance of magnetoelastic coupling for chiral magnets and shed light on uniaxial pressure control of the $\chi_s$ and topological Hall effect.

**Methods:**

**Sample Growth[14]:**

EuAl$_4$ crystals were grown from a high-temperature aluminum-rich melt. Eu pieces and Al shot totaling 2.5 g were loaded into one side of a 2-mL alumina Canfield crucible set. The crucible set was sealed under 1/3 atm argon in a fused silica ampoule. The ampoule assembly was placed in a box furnace and heated to 900°C over 6 h (150°C/h) and held for 12 h to melt and homogenize the metals. Crystals were precipitated from the melt during a slow cool to 700°C over 100 h (−2°C/h).

**XRMS:**

The XRMS measurements were performed at the POLAR beam line of the Advanced Photon Source (APS), Argonne National Laboratory (ANL). The photon energy was tuned to the Eu $L_2$ resonance (7.617 keV) using a double-crystal Si (111) monochromator. Circularly polarized x rays were generated using a 180-μm-thick diamond (111) phase plate, and focused to about 2.5 × 2.5 μm$^2$ full width at half maximum (FWHM) using Be compound refractive lenses. Temperature was controlled using a He flow cryostat coupled to a He recirculator. The diffracted signal was measured in reflection from the sample [001] surface using horizontal scattering geometry and an Eiger 1M area detector.

Both magnetic and structural peaks show a Full-Width-at-Half-Maximum (FWHM) better than 0.02°. Even small changes peak width of the sample can induce large variations in peak intensity. Therefore, for the acquisition of the XRMS images with the area detector, we measure integrated intensities at each point by rocking the sample by 0.2° while counting.

**XRMS cross-section:**

Under the electric dipole approximation, the resonant scattering process involves transitions between the core state $|\zeta_v\rangle$ with energy $E_v$ and an unoccupied state $|\psi_\eta\rangle$ with energy $E_\eta$ in both absorption and emission channel. The scattering amplitude can be written as:

$$f_{res}(\omega) = \sum_{ij} \hat{\epsilon}'_i \hat{\epsilon}_j \sum_\eta \frac{\langle \zeta_v|R_i|\psi_\eta\rangle\langle\psi_\eta|R_j|\zeta_v\rangle}{\omega-(E_\eta-E_v)+i\Gamma} = \sum_{ij} \hat{\epsilon}'_i \hat{\epsilon}_j T_{ij} \quad (M1)$$

where $\hat{\epsilon}$ and $\hat{\epsilon}'$ are the polarization vectors of the incident and scattering x-rays, respectively. $R$ is the position operator. When the resonant atom has a parity-even magnetic moment $\widehat{M}_n$ at site $n$, and assuming cylindrical symmetry, the magnetic scattering takes the form:[27]

$$f_n(\omega) = (\hat{\epsilon}'\cdot\hat{\epsilon})f_0(\omega) - i(\hat{\epsilon}'\times\hat{\epsilon})\cdot\widehat{M}_n f_1(\omega) + (\hat{\epsilon}'\cdot\widehat{M}_n)(\hat{\epsilon}\cdot\widehat{M}_n)f_2(\omega) \quad (M2)$$

Since $f_2(\omega)$ is usually much smaller than $f_1(\omega)$, the magnetic scattering is dominated by the second term of Eq. (M2). For the 1D helical SDW, the CD has been derived respectively for the tilting and rocking geometries[28]. Here rocking and tilting geometries refer to sample rotations along an axis that is perpendicular and parallel to the scattering plane, respectively. As an example, the rocking and tilting geometries in Fig. 1**a** are corresponding sample rotations along the y- and x-axis, respectively. For the chiral Bloch type SDW:

$$I^{CR}(\boldsymbol{Q}^{\pm}_\chi) - I^{CL}(\boldsymbol{Q}^{\pm}_\chi) = (\tau\chi_s)\mathcal{D}^{xz}(\boldsymbol{Q}^{\pm}_\chi) \quad \text{in the rocking geometry,} \quad (M3)$$

$$I^{CR}(\boldsymbol{Q}^{\pm}_\chi) - I^{CL}(\boldsymbol{Q}^{\pm}_\chi) = \chi\mathcal{B}^{yz}(\boldsymbol{Q}^{\pm}_\chi) \quad \text{in the tilting geometry,} \quad (M4)$$

where $\tau=sign((\boldsymbol{Q}^{\pm}_\chi \cdot \vec{x})/|\boldsymbol{Q}^{\pm}_\chi \cdot \vec{x}|)$. The $\vec{x}$ is the unit vector along the x-direction (Fig. 1**a**) and $\boldsymbol{Q}^{\pm}_\chi = \boldsymbol{q}^{\pm}_\chi + \boldsymbol{G}$ is the magnetic superlattice peak position in Brillouin zone $\boldsymbol{G}$. The $\mathcal{D}^{xz}$ and $\mathcal{B}^{yz}$ is a function of magnetic moment that is independent of $\tau$ and $\chi_s$. The superscripts highlight the sample-rotation planes. Note that in the tilting geometry the CD will not change sign from $\boldsymbol{Q}_\chi$ to $-\boldsymbol{Q}_\chi$. Our experiment was performed in the rocking geometry that shows sign change for the chiral Bloch type SDW.

**Acknowledgements:** We thank Yue Cao, Miao-Fang Chi, Jong-Woo Kim, Kaifa Luo, Michael McGuire, Philip Ryan, and Jiaqiang Yan for stimulating discussions. This research was supported by the U.S. Department of Energy (DOE), Office of Science, Basic Energy Sciences, Materials Sciences and Engineering Division (part of x-ray measurement and data analysis, and sample growth). This research is also sponsored by the Laboratory Directed Research and Development Program of Oak Ridge National Laboratory, managed by UT-Battelle, LLC, for the US Department of Energy (part of x-ray measurement and data analysis). CD-XRMS was performed on beam time awards (DOIs: 10.46936/APS-190778/60014621, 10.46936/APS-189209/60013667) from the Advanced Photon Source, a U.S. DOE Office of Science user facility at Argonne National Laboratory, and is based on research supported by the U.S. DOE Office of Science-Basic Energy Sciences, under Contract No. DE-AC02-06CH11357.

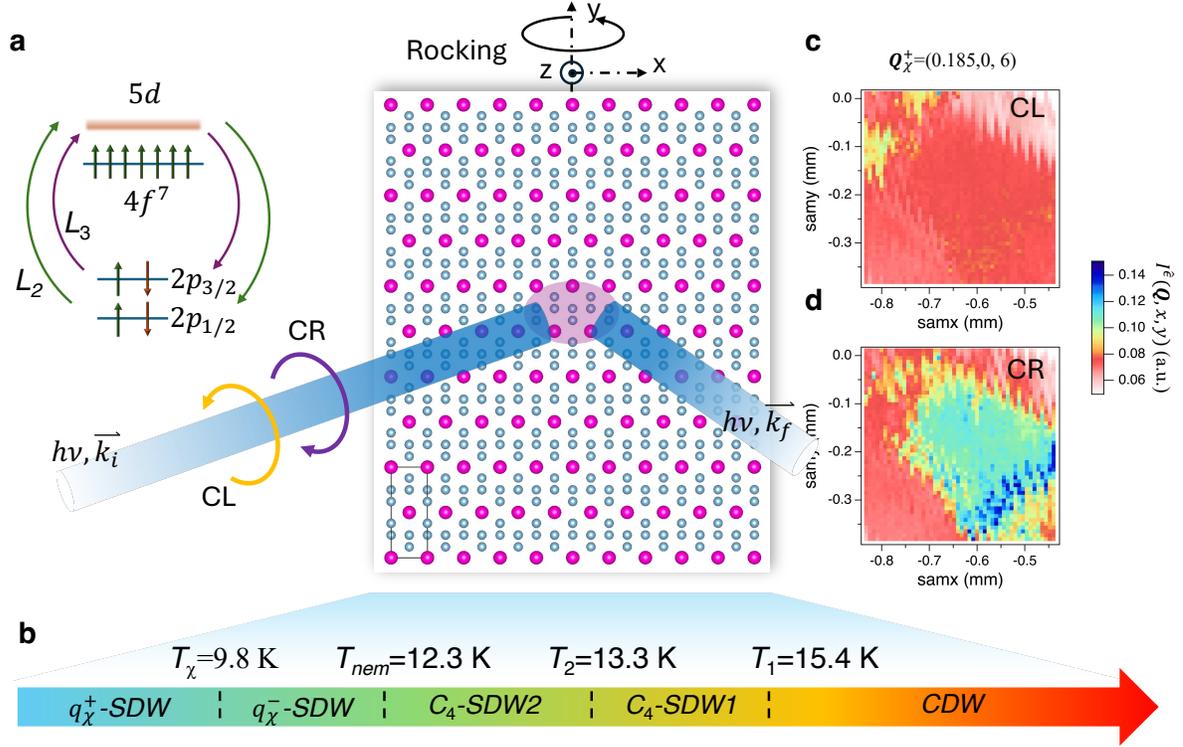

**Figure 1: XRMS imaging of spin, charge, orbital, and lattice structures. a,** schematics of XRMS imaging. The incident photon with energy, $h\nu$, momentum, $\mathbf{k}_i$, and polarization, $\hat{\epsilon}$, is scattered to final state ($h\nu$, $\mathbf{k}_i$, $\hat{\epsilon}'$). By tuning $h\nu$ to the Eu $L_{2,3}$-edges, the core electrons at the $2p_{1/2,3/2}$ levels are excited to the unoccupied $5d$ state. Combined with control of incident polarization and high spatial resolution, the spin, charge, orbital, and lattice orders can be selectively imaged in the same FOV. The rocking and tilting geometry refer to sample rotations along an axis that is perpendicular and parallel to the scattering plane, respectively. In this work, we use Eu $L_2$-edges XRMS to image the magnetic domains of EuAl$_4$. **b,** cascade symmetry breakings of EuAl$_4$ at zero external magnetic field. The CDW induces transverse lattice deformation that breaks inversion and mirror-symmetries. **c** and **d,** prototypical XRMS images $I^{CL}(\mathbf{Q},x,y)$ and $I^{CR}(\mathbf{Q},x,y)$. Images were collected at $T=8.5$ K in a 400×400 μm² area. The vertical and horizontal optical beam size was set to 2.5 μm. The diffraction condition was set to $\mathbf{Q} = \mathbf{Q}_{sdw} = (0.185, 0, 6)$ to selectively probe the chiral SDW domain. The values of all XRMS images shown in this study was obtained by counting the intensity for 2 second while rocking the sample in 0.2° (see Methods).

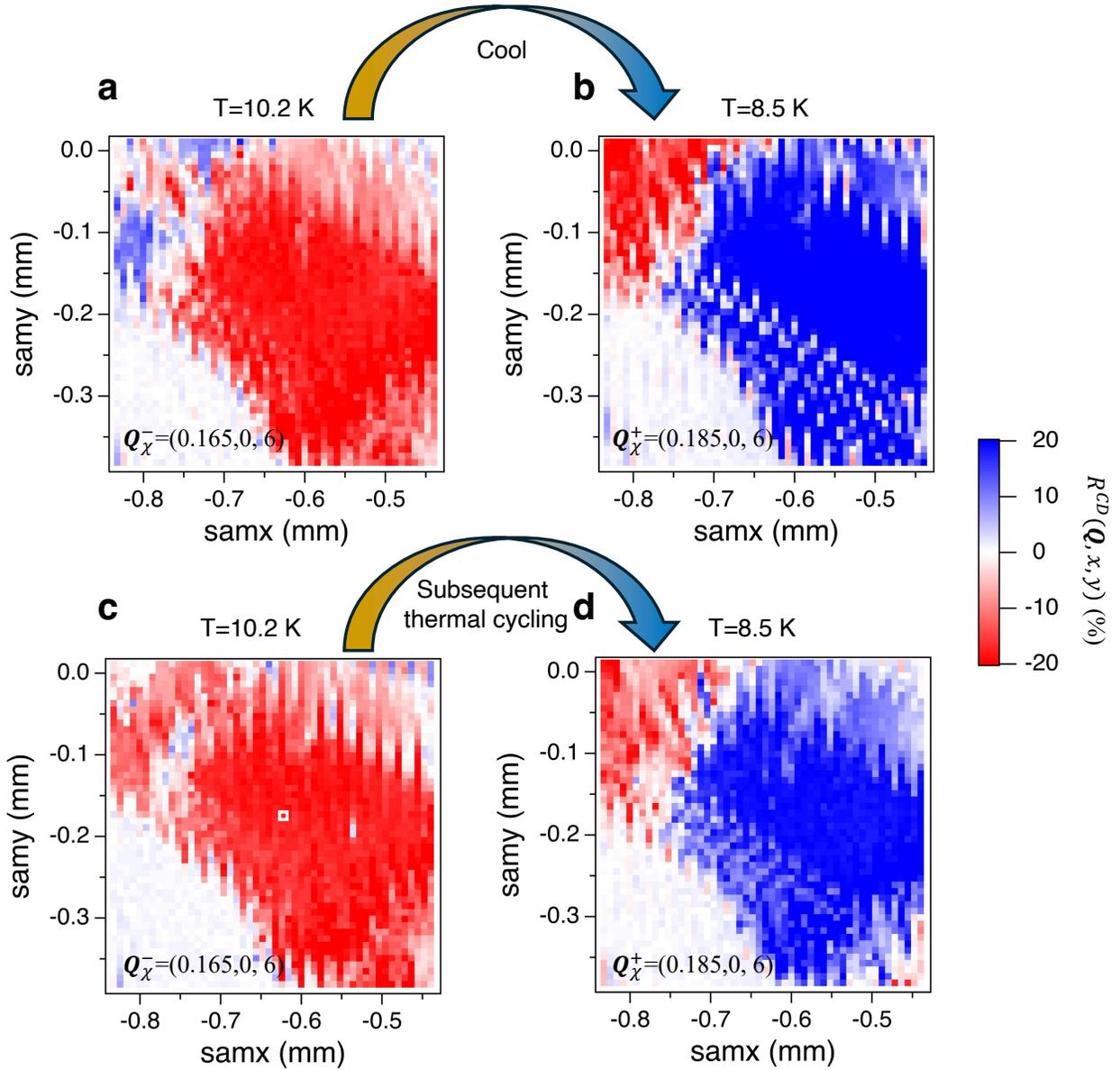

**Figure 2: Macroscopic chiral memory effect. a** and **b** show $I^{CD}(\mathbf{Q}_{sdw}, x, y)$ above (10.2 K) and below (8.5 K) $T_\chi$. The entire 200 μm-scale $\chi$=-1 magnetic domain changes to $\chi$=+1 below $T_\chi$. **c** and **d** show the same $I^{CD}(\mathbf{Q}_{sdw}, x, y)$ at 10.2 and 8.5 K as **a** and **b**, but are measured after a thermal cycling to 20 K. Both the chirality and size of the main chiral magnetic domain remain nearly unchanged, representing a macroscopic chiral memory effect. Note after the thermal cycling, the chirality of the top left domain is pinned to negative, due to larger lattice defects in that area (see Supplementary Fig. S2).

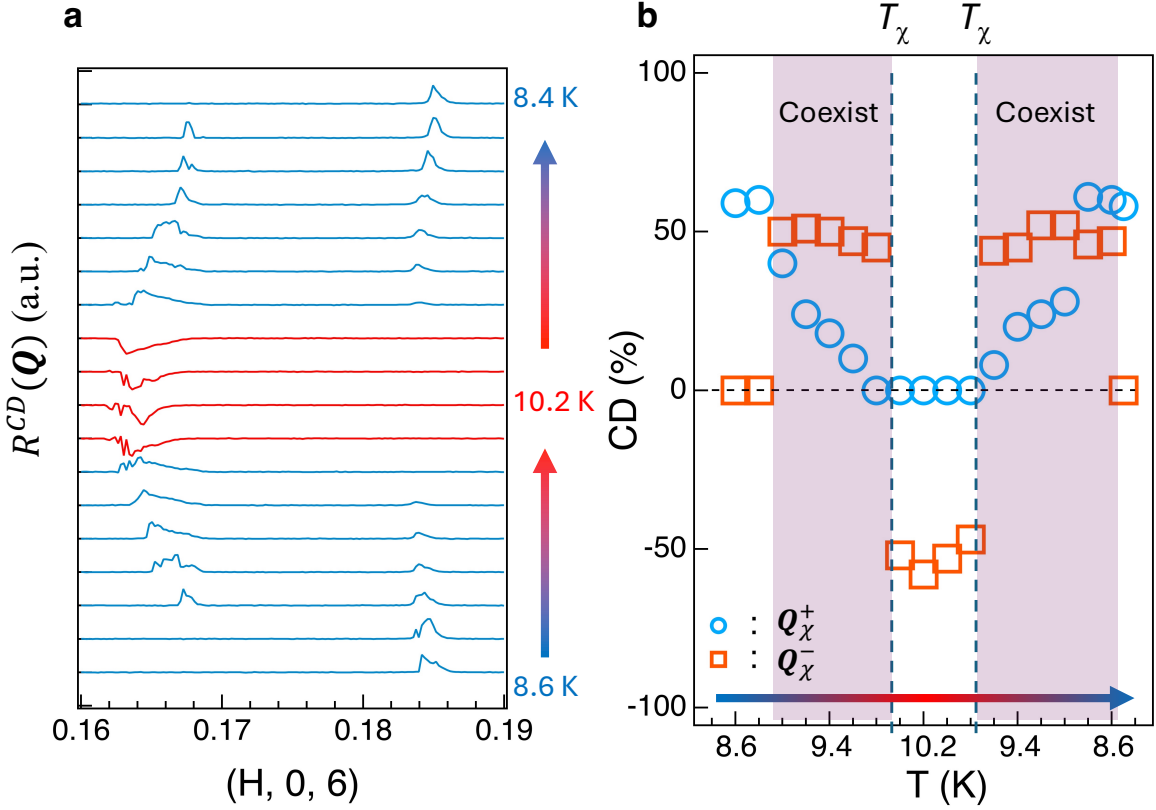

**Fig. 3: The coupling between $\chi_s$ and $q_{SDW}$. a,** Temperature dependence of the CD-XRMS line scan along the (H, 0, 6) direction. The spot size of the incident beam was set to 2.5 × 2.5 μm² for these line scans. These scans were measured at white square position shown in Fig. **2c**. We first warm up from 8.6 to 10.2 K and then cool down to 8.4 K. The temperature step is 0.2 K. Cyan curves highlight scans that show the magnetic superlattice peak at $Q_\chi^+$. Red curves represent scans that show the magnetic superlattice peak only at $Q_\chi^-$. The phase coexistence temperature is about 1 K with a hysteresis of ~0.3 K. The extracted peak intensities of the CD scans in **a** are shown in **b**. The cyan and red marks represent $Q_\chi^+$ and $Q_\chi^-$, respectively.

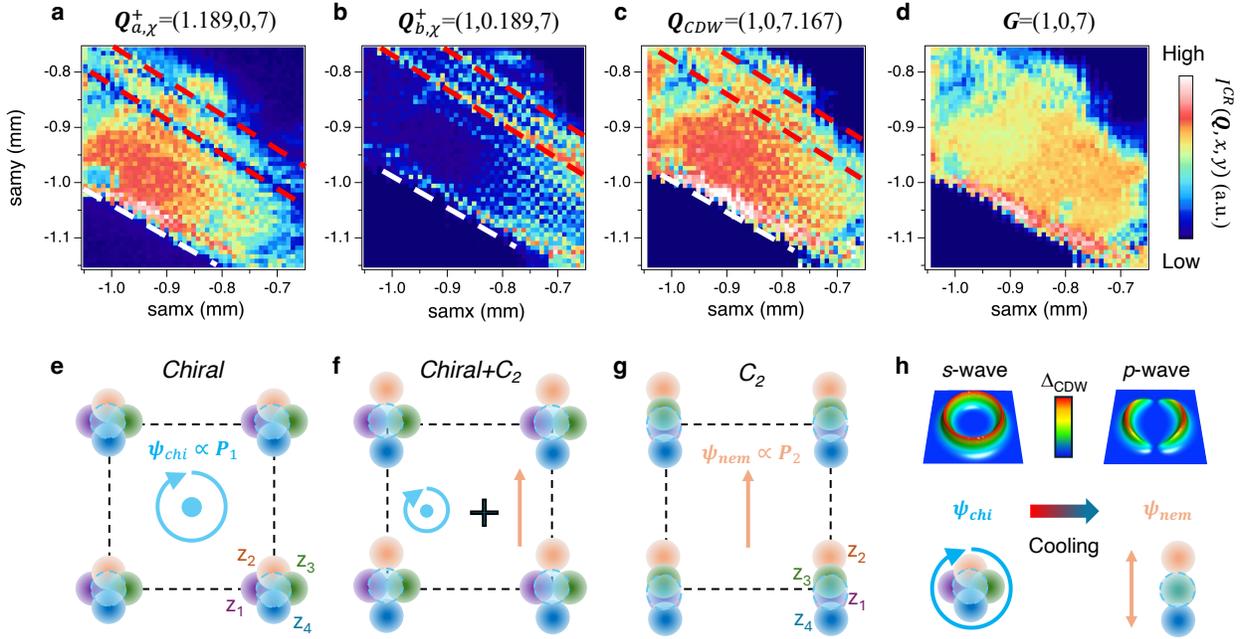

**Figure 4: CDW pinning of chiral magnetic domain. a-d,** $I^{CR}(Q, x, y)$ at $Q=Q^+_{a,\chi}=(1.189, 0, 7)$, $Q^+_{b,\chi}=(1, 0.189, 7)$, $Q_{CDW}=(1, 0, 7.17)$, and $G=(1, 0, 7)$ at $T=5.5$ K. We note that $Q^+_\chi$ displays temperature dependence and changes from $q^+_\chi=(0.185,0,0)$ at 9 K to $(0.189, 0, 0)$ at 5.5 K.[15] **a** and **b** are corresponding to orthogonal SDW domains with spin modulation along the b-axis and a-axis, respectively. **c** and **d** select incommensurate-CDW and structural domains. The measurements were performed on the same sample as in Figs. 1-3, but during a separate beamtime. Consequently, the FOV is different from that shown in Figs. 1-3. The white dashed line marks the sample edge. The red dashed lines highlight the intensity gaps in **a** and **c**, and intensity peaks in **b**. **e-g,** schematic lattice deformations in the paramagnetic and chiral CDW (C4), mixed chiral and nematic CDW (C2), and pure nematic CDW phases (C2).[15] The incommensurate nematic CDW induces in-plane polarization, $P_2$, that are different from the chiral CDW with out-of-plane polarization $P_1$. The color of the solid circles represents different atomic height ($z_{1,2,3,4}$) along the crystalline c-axis. Dashed circles represent the undistorted atomic position in the ab-plane. **h,** competing chiral and nematic CDWs. The chiral CDW favors C4 symmetry and *s*-wave CDW gap in the ab-plane, whereas nematic CDW breaks C4 symmetry and host *p*-wave-like CDW gap in the ab-plane. The spin chirality flips when $\psi_{chi}$ is suppressed by $\psi_{nem}$. In this picture, the $q^-_\chi$ phase is a metastable phase for which $\psi_{chi}$ and $\psi_{nem}$ microscopically coexist.